\documentclass[11pt,cite,epsfig,psfrag]{article}
\usepackage[utf8]{inputenc}
\usepackage{placeins}
\usepackage{geometry}
 \usepackage{graphicx}
 \usepackage{subfig}
  \usepackage{amsmath} 
  \graphicspath{{images@4dGBBI/}}
  \usepackage[usenames, dvipsnames]{color}
\usepackage{wrapfig}
\usepackage{boxedminipage}
\usepackage{multirow}

\usepackage{fullpage}
\usepackage{amsmath}
\usepackage{amssymb}
\usepackage{graphicx}
\usepackage{mathrsfs}
\usepackage{wrapfig}
\usepackage{boxedminipage}
\usepackage{epsfig}
\usepackage{setspace}
\usepackage{float}
\usepackage{hyperref}
\usepackage{enumerate}
\usepackage{cite}
\usepackage[font=footnotesize,labelfont=rm]{caption}

\usepackage[numbers,sort&compress]{natbib}

\def\be{\begin{equation}}
\def\ee{\end{equation}}
\def\bea{\begin{eqnarray}}
\def\eea{\end{eqnarray}}

\numberwithin{equation}{section}

 \newcommand{\RN}[1]{%
   \textup{\uppercase\expandafter{\romannumeral#1}}%
 }
\begin{document}

\thispagestyle{empty}
%\baselineskip 20pt
%\rightline{IP/BBSR/2005-12}
%\rightline{\tt hep-th/yymmnnn}

\vskip 2cm

\begin{center}
{\Large \bf Topology of  Born-Infeld AdS black holes in 4D novel Einstein-Gauss-Bonnet gravity}
\end{center}

\vskip .2cm

\vskip 1.2cm

\centerline{ \bf   Pavan Kumar Yerra \footnote{pk11@iitbbs.ac.in} and Chandrasekhar Bhamidipati\footnote{chandrasekhar@iitbbs.ac.in}
}

\vskip 7mm 
\begin{center}{ $^{1}$ Institute of Physics, Sachivalaya Marg, \\ Bhubaneswar, Odisha, 751005, India}	
\end{center}

\begin{center}{ $^{2}$ School of Basic Sciences\\ 
Indian Institute of Technology Bhubaneswar \\ Bhubaneswar, Odisha, 752050, India}
\end{center}

\vskip 1.2cm
\vskip 1.2cm
\centerline{\bf Abstract}
\vskip 0.5cm
\noindent
The topological classification of critical points of black holes in 4D Einstein-Gauss-Bonnet gravity coupled to Born-Infeld theory is investigated. Considered independently, Born-infeld corrections to the Einstein action alter the topological charge of critical points of the charged AdS black hole system, whereas the Gauss-Bonnet corrections do not. For the combined system though, the topological charge of the Einstein-Gauss-Bonnet theory is unaltered in the presence of Born-Infeld coupling.

\newpage
\setcounter{footnote}{0}
\noindent

\baselineskip 15pt

%%%%%%%%%%%%%%%%%%%%%%%%
\section{Introduction}
%%%%%%%%%%%%%%%%%%%%%%%%

Black holes offer a unique window to investigate the effects of strong gravitational phenomena~\cite{LIGO2017}, and an understanding of their thermodynamics~\cite{Bekenstein:1973ur,Bardeen:1973gs,Hawking:1975vcx} is perhaps the key to insights in to the nature of degrees of freedom of gravity~\cite{Sen:2007qy}. Phase transitions of black holes, in particular the Hawking-Page transition~\cite{Hawking:1982dh}, evoke compelling interest, in part due to possible interpretation in the context of gauge/gravity duality~\cite{,Maldacena:1997re,Gubser:1998bc,Witten:1998qj}. For charged black holes in anti de Sitter (AdS) spacetimes, a first-order phase transition takes place between small and large black holes, very similar to the liquid-gas transitions which occur in van der Waals (VdW) fluid~\cite{Chamblin:1999tk}. These investigations become more invigorating in the black hole chemistry program~\cite{Caldarelli:1999xj,Kastor:2009wy,Cvetic:2010jb,Dolan:2011xt,Karch:2015rpa,Kubiznak:2016qmn}, where the cosmological constant $\Lambda$ is treated as a variable, leading to the concept of a pressure $P = -\Lambda/8\pi$, with a thermodynamic volume $V$ acting as its conjugate. In this novel set up, the first law $dM = T dS $ assumes the form
$dM = T dS + VdP \, $, where $M$ is now reinterpreted as enthalpy $H$~\cite{Kastor:2009wy}. The phase transitions acquire an appropriate meaning and analogy with liquid-gas transitions is now complete, including the existence of a critical region where the first order phase transitions end in a second order transition~\cite{Chamblin:1999tk,Caldarelli:1999xj,Kubiznak:2012wp}.  \\

\noindent
Criticality in phase transitions in statistical physics and especially in black holes is an interesting topic on its own right due to emergence of several universal phenomena and scaling. In addition to the small to large black hole phase transition, there are now new examples of transitions such as reentrant transitions, solid-liquid-gas type transitions with multiple critical points in higher dimensions~\cite{Altamirano}. For many of these transitions to exist, one typically requires the presence of both gauge and gravity higher curvature corrections such as Born-Infeld (BI) and Gauss-Bonnet (GB) terms added to the Einstein action or involving Kerr-AdS black holes with multiple rotation parameters. In fact, the later system can also exhibit tricritical points though in a restricted parameter range, including the existence of superfluidity~\cite{Lu:2010xt,Shen:2005nu,Cvetic:2001bk,Cai:2001dz,AltamiranoKubiznak,Altamirano,Altamirano:2013ane,Cai:2013qga,Kofinas:2006hr,Wei:2019uqg,Hennigar:2016xwd,Banerjee:2011cz,Pourhassan:2020bzu}. 
Thermodynamics of black holes in the presence of higher derivative curvature terms of gauge and gravity is important from holographic view point too, as  these appear naturally in discussions involving semiclassical quantum gravity and also in the low energy limit of certain superstring theories\cite{Boulware:1985wk,Cai:2001dz}. In the context of AdS/CFT correspondence, Gauss-Bonnet terms for example are useful as they lead to corrections in the large $N$ expansion of gauge theories on the boundary, in the strong coupling limit. Such terms have in fact given important contributions to the viscosity to entropy ratio, apart from leading to novel bounds~\cite{Brigante:2007nu}.  A convenient advantage involving Gauss-Bonnet and Lovelock terms in general is that the field equations in these theories do contain more than two derivatives of the metric in five and higher dimensions, are topological in four dimensions, and vanish for dimension less than three. The general Lovelock terms have similar characteristics in their critical dimension
and are useful due to the absence of ghost-like modes.\\

\noindent 
It is then interesting that in a novel limit obtained by rescaling the Gauss-Bonnet coupling as $\alpha \to \alpha/(D-4)$ in four dimensions \cite{Tomozawa:2011gp,Cognola:2013fva}, non-trivial theory could be obtained, giving the new 4D Einstein Gauss-Bonnet (EGB) theory~\cite{Glavan:2019inb}. There have been a lot of developments with black hole solutions obtained in a with a wide variety of non-trivial coupling to matter and gauge terms in this novel gravity theory~\cite{Fernandes:2020rpa,Kumar:2020xvu,Hegde:2020yrd,Wei:2020poh,HosseiniMansoori:2020yfj,Ghosh:2020ijh,Wei:2014hba,Wei:2012ui,Frassino:2014pha,Akbarieh:2021vhv,Ghosh:2020syx,Jaryal:2022rzd}. In particular, Born-Infeld coupled Einstein gravity is important to study as a possible regular theory over the Maxwell electrodynamics, as the later usually includes infinite self energy due to a singularity at the position of charge\cite{Born:1934gh,Hoffmann:1935ty}. Further, the Born-Infeld terms are also appear from the low energy effective action of open string theories. In the past, black hole solutions in these theories have garnered wide attention\cite{Dey:2004yt,Cai:2004eh}, which has become more interesting in the extended thermodynamic framework, as these theories exhibit novel reentrant phase transitions in four dimensional Einstein Born-Infeld gravity, apart from the existence of the small and large black hole transition, similar to the vdW system\cite{Gunasekaran:2012dq,Zou:2013owa}. In addition to the traditional treatments of phase transitions in black holes which giving useful information, it is important to look for novel methods to classify the critical points in general black hole systems.\\

\noindent
One new idea involves a topological approach to classification of critical points of phase transitions in general theories of gravity, proposed in~\cite{Wei:2021vdx}, where each critical point can be bestowed with a topological charge. For instance, critical points which carry a topological charge $Q_t=-1$ are termed conventional, where as the ones with opposite topological charge are called novel. For consistency in situations where there are multiple critical points, a more appropriate classification is to think of the former (later) type as coming from disappearance (appearance) of new phases~\cite{Yerra:2022alz}, which seems to hold in the case of isolated critical points as well~\cite{Ahmed:2022kyv}. An unexpected result is that, the Born-Infeld corrections to the action change the topological class of critical points of black holes~\cite{Wei:2021vdx}, where as the Gauss-Bonnet corrections do not~\cite{Yerra:2022alz}. To get deeper insight into this intriguing topological behavior, it is desirable to study nature of critical points in a more general theory of gravity, possibly by exploring black holes in a system consisting of both types of terms. The main objective of this letter is to report progress in this regard, by investigating the topological nature of critical points in a black hole solution in 4D Einstein Gauss-Bonnet (EGB) gravity coupled to Born-Infeld Electrodynamics\cite{Born:1934gh,Hoffmann:1935ty,Dey:2004yt,Cai:2004eh,Gunasekaran:2012dq,Zou:2013owa} found in~\cite{Yang:2020jno}. The thermodynamics and phase transitions in the extended thermodynamic framework has been explored earlier in~\cite{Zhang:2020obn}.\\

\noindent
The rest of the paper is organised as follows. In section-(\ref{phasestructure}), we closely follow~\cite{Yang:2020jno,Zhang:2020obn,Wei:2014hba,Duan:2018rbd,Duan:1984ws} and collect salient aspects of thermodynamics of 4D Einstein Gauss-Bonnet gravity coupled to Born-Infeld AdS black holes in four dimensions and briefly present the topological approach required for analysis in section(\ref{4GBBI}). Our main calculations and results are presented in section-(\ref{4GBBI}) and the discussion of the topological nature of critical  points is given in subsection-(\ref{naturegbbi}). Remarks and conclusions are given in section-(\ref{conclusions}).

%%%%%%%%%%%%%%%%%%%%%%%%%%%%%%%%%%%
\section{Thermodynamic phase structure and Topology} \label{phasestructure}
%%%%%%%%%%%%%%%%%%%%%%%%%%%%%%%%%%%%%%%%
We start with the action of  $D$-dimensional Einstein-Gauss-Bonnet (EGB) gravity
	minimally coupled to the Born-Infeld (BI) electrodynamics in the presence of  a negative cosmological constant $\Lambda~\footnote{ $\Lambda \equiv -\frac{(D-1)(D-2)}{ 2l^2}$, and $l$ is the AdS length. }$, given by~\cite{Yang:2020jno,Zhang:2020obn}:
	%\frac{\alpha}{D-4}
	\begin{equation}\label{action}
	\mathcal{I} = \frac{1}{16\pi} \int d^D x \sqrt{-g} \Big(R-2\Lambda + \alpha \mathcal{G} + \mathcal{L_{BI}} \Big),
	\end{equation} where $\mathcal{G} = R^2 -4R_{\mu \nu} R^{\mu \nu} + R_{\mu \nu \rho \sigma} R^{\mu \nu \rho \sigma} $ is the GB term, $\alpha > 0$ is the GB coefficient, and the BI Lagrangian with coupling parameter $\beta >0$, is
	 \begin{equation}
	\mathcal{L_{BI}}= 4\beta^2 \bigg(1- \sqrt{1+\frac{F_{\mu \nu} F^{\mu \nu}}{2\beta^2}} \bigg).
	\end{equation} 
In $D=4$, the Gauss-Bonnet term becomes a topological invariant and does not contribute to the equations of motion. In $D>4$, the the above theory is known to admit analytical static spherically symmetric solutions together with the Born-Infeld term, with a well defined thermodynamics. More recently,  following the insightful proposal of Glavan and Lin~\cite{Glavan:2019inb} and replacing, 
\begin{equation}
\label{limit}
\alpha\rightarrow \frac{\alpha}{D-4}\,,
\end{equation}
in the action in eqn. (\ref{action}), and then
followed by the limit $D\rightarrow 4$ still retains non-trivial effects of the Gauss-Bonnet term on various properties of black hole solutions. In the $D \rightarrow 4 $ limit, the D-dimensional spherically symmetric solution of the above action admits Born-Infeld AdS black holes in 4-dimensional Einstein-Gauss-Bonnet gravity with the following metric~\cite{Yang:2020jno}: 
	\begin{equation}\label{eq:metric}
	ds^2  =  -f(r) dt^2 + \frac{dr^2}{f(r)} + r^2(d\theta^2 + \text{sin}\theta ^2 d\varphi ^2),
	\end{equation}
	where
	\begin{equation} \label{eq:f(r)}
	 f(r) = 1+\frac{r^2}{2\alpha} \Big\{ 1-\bigg(1+4\alpha \Big(  \frac{2M}{r^3}-\frac{1}{l^2} -\frac{2\beta^2}{3}+\frac{2\beta^2}{3} \sqrt{1+ \frac{Q^2}{\beta^2 r^4}} -\frac{4Q^2}{3r^4} \text{}_{2}F_{1} \Big[\frac{1}{4}, \frac{1}{2}, \frac{5}{4}, \frac{-Q^2}{\beta^2 r^4}\Big]  \Big)  \bigg)^{\frac{1}{2}}  \Big\}.
	\end{equation} 
Here,  $\text{}_{2}F_{1}$ is the hypergeometric function, $M$ is the mass and $Q$ is the charge of the black hole.
We note here that, in the limit of $\alpha \rightarrow 0$, the above thermodynamic quantities correspond to  Born-Infeld AdS black holes~\cite{Gunasekaran:2012dq}, whereas, in the limit of $\beta \rightarrow \infty$, they correspond to charged AdS black holes in four dimensional novel Einstein-Gauss-Bonnet gravity~\cite{Wei:2020poh,Fernandes:2020rpa}.
\vskip 0.3cm
\noindent
However, it has been pointed out that the aforementioned $D \rightarrow 4$ limit leading to 4D EGB theory as originally proposed by Glavan and Lin~\cite{Glavan:2019inb} is actually inconsistent~\cite{Arrechea:2020evj,Ai:2020peo,Mahapatra:2020rds,Shu:2020cjw} and resulting thermodynamics is ill defined. It does not admit a description in terms of a covariantly-conserved rank-2 tensor in four dimensions and	the dimensional regularization procedure needs to be clarified~\cite{Gurses:2020ofy}. The theory probably makes sensible only in some  highly-symmetric spacetimes ( such as the FLRW  and static spherically symmetric spacetimes)~\cite{Yang:2020jno}. Recently, certain Kaluza-Klein dimensional reduction and counter-term methods have been used to formulate a consistent 4D EGB gravity, where  the resulting theories now contain an extra scalar degree of freedom~\cite{Fernandes:2020nbq,Hennigar:2020lsl,Kobayashi:2020wqy,Lu:2020iav}. The black hole solution used in this work though (eqs. \ref{eq:metric} and \ref{eq:f(r)}), is also a solution of this consistent 4D EGB gravity (which is a special scalar-tensor theory that belongs to the family of Horndeski gravity), obtained by compactifying D-dimensional EGB gravity on a $(D-4)$-dimensional maximally symmetric space~\cite{Lu:2020iav,Yang:2020jno}. The metric profile is independent of the curvature of the internal space  and  the scalar field. Thus, with these limitations, we proceed to obtain the thermodynamic quantities (seen below in eqs.~\ref{eq: temp}~to~\ref{eq: gibbs}) including the entropy, which follow from this consistent theory~\cite{Lu:2020iav,Yang:2020jno}. \\
	
\noindent	
In the extended phase space (where we can have pressure $P$, via $P=-\Lambda/8\pi$, and its conjugate thermodynamic volume $V$~\cite{Kastor:2009wy}),   the thermodynamic quantities, temperature $T$, entropy $S$, electromagnetic potential $\Phi_{em}$, mass $M$, and Gibbs free energy $G$ of these black holes are  given (in terms of horizon radius $r_+$) by~\cite{Yang:2020jno,Zhang:2020obn}:
\begin{eqnarray}
T &=& \frac{r_+^2-\alpha + 8\pi P r_+^4 + 2r_+^4\beta^2 \Big( 1- \sqrt{1+ \frac{Q^2}{r_+^4\beta^2}}\Big)}{4\pi r_+ (r_+^2+2\alpha)}, \label{eq: temp} \\
S &=& \pi r_+^2 +4\pi \alpha \text{ln} \frac{r_+}{\alpha}, \label{eq: entropy} \\
\Phi_{em} &=& \frac{Q}{r_+} \text{}_{2}F_{1} \Big[\frac{1}{4}, \frac{1}{2}, \frac{5}{4}, \frac{-Q^2}{\beta^2 r_+^4}\Big], \\
M &=& \frac{r_+}{2} \bigg[1+ \frac{8\pi P r_+^2}{3} +\frac{\alpha}{r_+^2} + \frac{2 \beta^2 r_+^2}{3} \bigg(1-\sqrt{1+ \frac{Q^2}{\beta^2 r_+^4}} \bigg) +\frac{4Q^2}{3r_+^2} \text{}_{2}F_{1} \Big[\frac{1}{4}, \frac{1}{2}, \frac{5}{4}, \frac{-Q^2}{\beta^2 r_+^4}\Big] \bigg], \, \, \\
G &=& M-TS. \label{eq: gibbs}
\end{eqnarray}

%%%%%%%%%%%%%%%%%%%%%%%%%%
\subsection{Phase Structure }
%%%%%%%%%%%%%%%%%%%%%%%%%%%
\begin{figure}[h!]
	% \begin{wrapfigure}{r}{0.43\textwidth}
	%\begin{center}
	{\centering
		\subfloat[]{\includegraphics[width=2in]{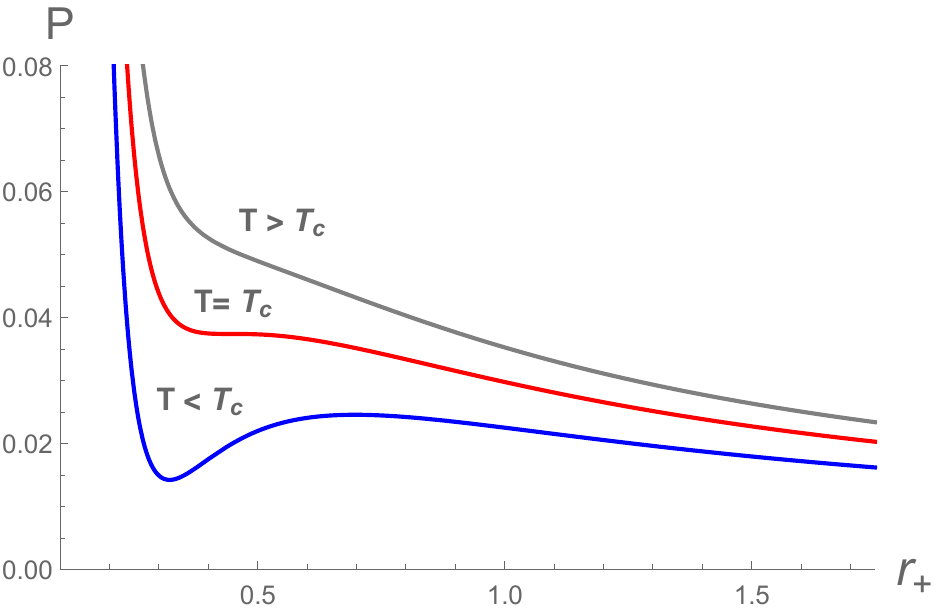}\label{Fig:pr1}}\hspace{0.2cm}	
		\subfloat[]{\includegraphics[width=2in]{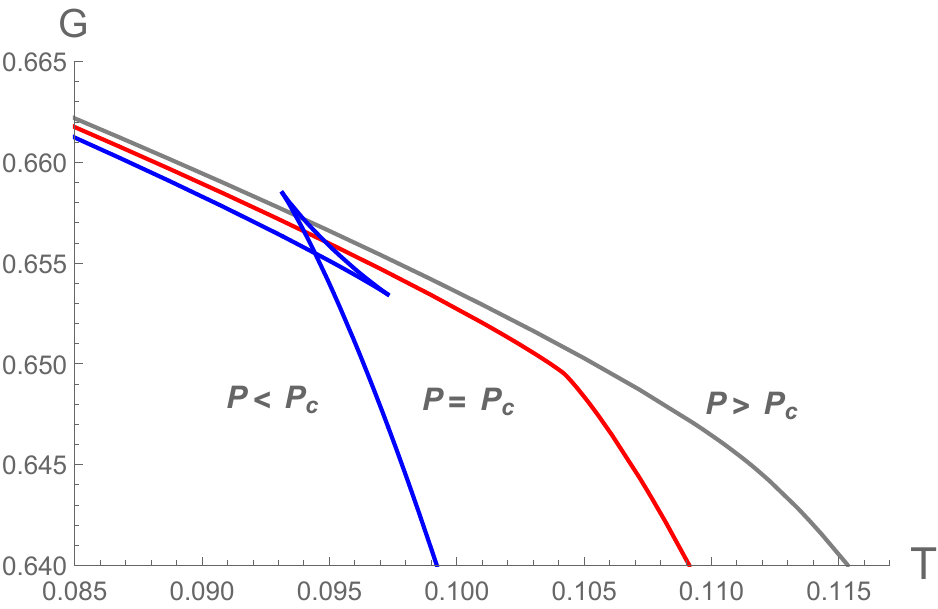}\label{Fig:gt1}}\hspace{0.2cm}				
    	\subfloat[]{\includegraphics[width=1.9in]{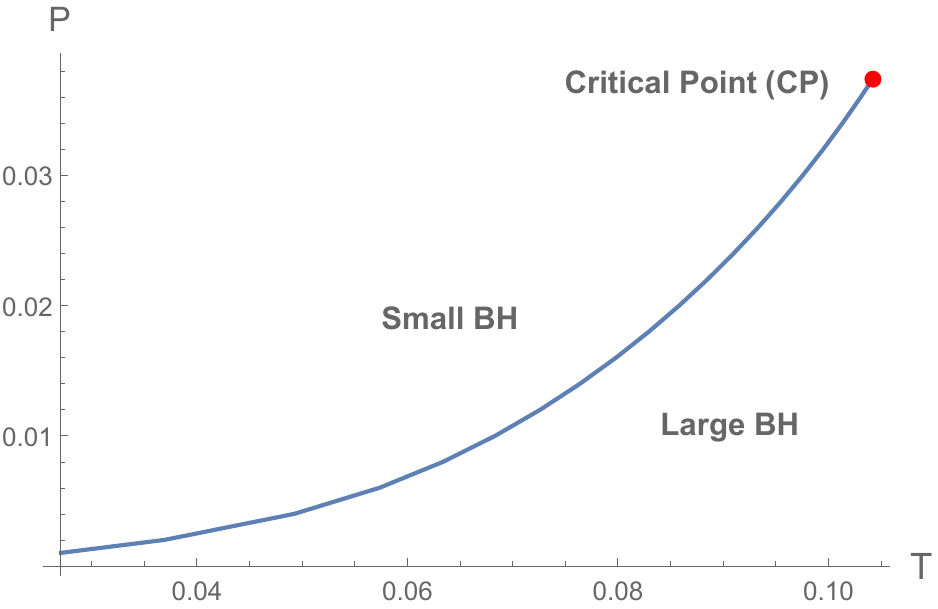}\label{Fig:pt1}}

		\caption{\footnotesize  (a) The behavior of the equation of state $P(r_+, T)$. (b) The behavior of the Gibbs free energy $G(T,P)$. (c) The coexistence line of small and large black hole phases. Plots are displayed for $(Q, \alpha, \beta)=(1,0.01,0.26)$. } 
	}
\end{figure} 

\begin{figure}[h!]
	% \begin{wrapfigure}{r}{0.43\textwidth}
	%\begin{center}
	{\centering
		\subfloat[]{\includegraphics[width=2.8in]{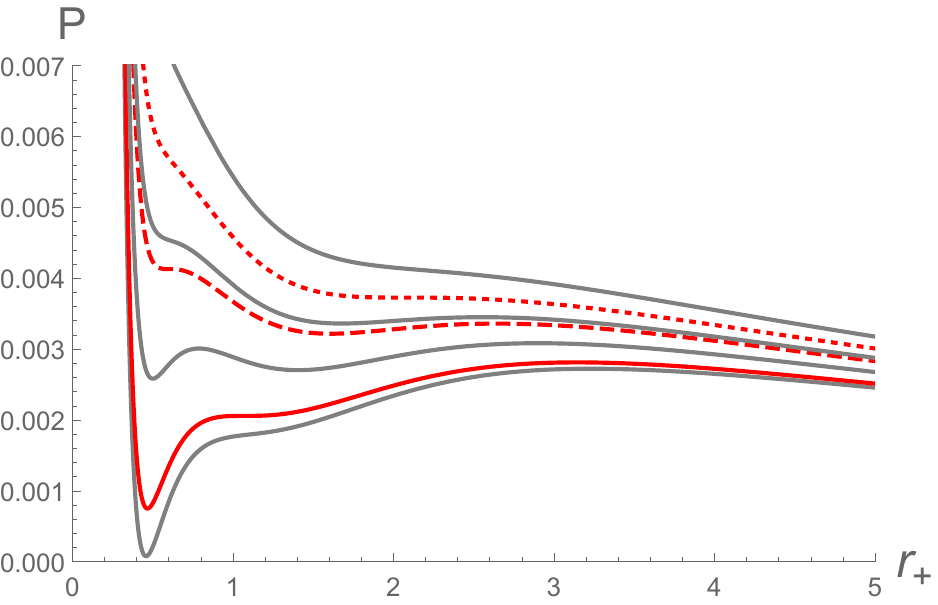}\label{Fig:pr3}}\hspace{0.5cm}	
		\subfloat[]{\includegraphics[width=2.8in]{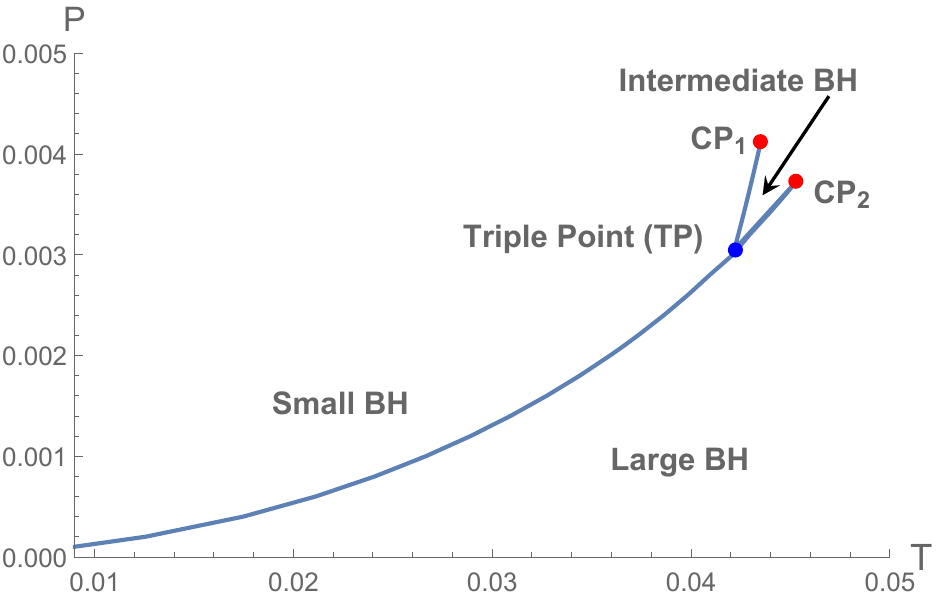}\label{Fig:pt3}}\hspace{0.5cm}				
		\subfloat[]{\includegraphics[width=2.8in]{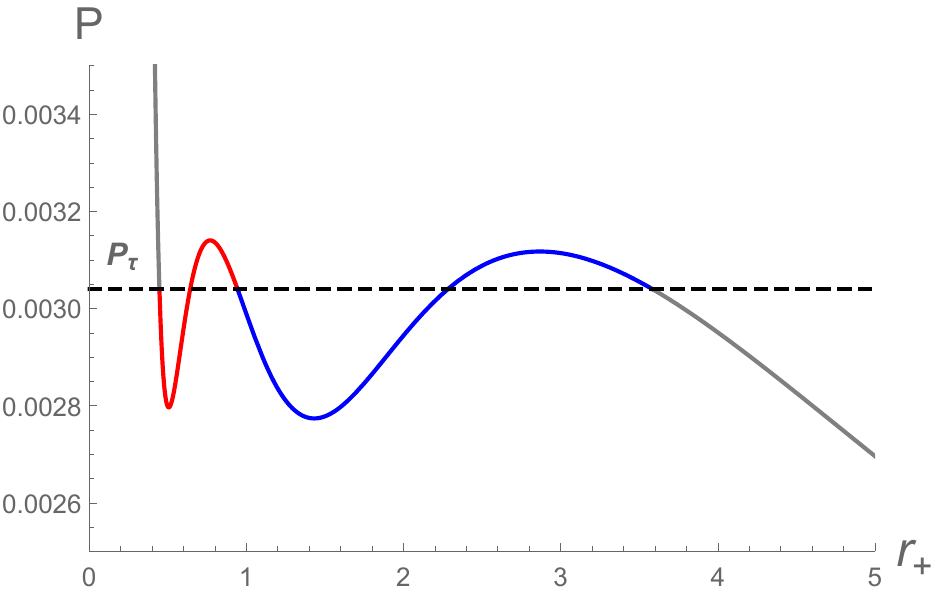}\label{Fig:prdst}}\hspace{0.5cm}
		\subfloat[]{\includegraphics[width=2.8in]{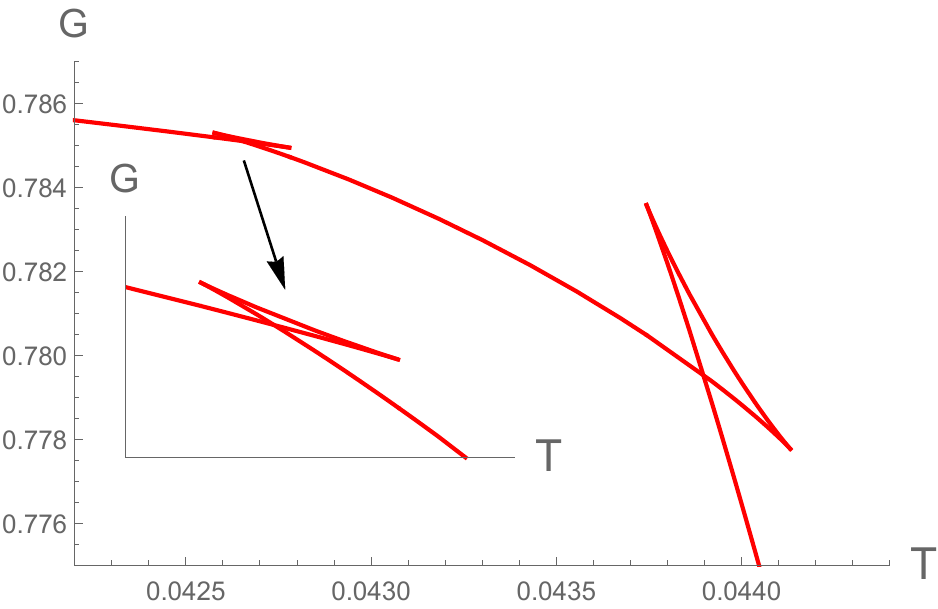}\label{Fig:gtdst}}

		\caption{\footnotesize  (a) The $P-r_+$ plot for different isotherms (temperature increases from bottom to top). Critical isotherms are shown with red color. (b) The $P-T$ plot shows the phase structure of small/intermediate/large black hole transitions and the triple point. (c) A special isotherm $(T=T_\tau)$, for which a ‘double’ Maxwell equal area holds for the same pressure $P_\tau$, gives the triple point $(P_\tau, T_\tau)=(0.00304, 0.0422)$. (d) The double swallow tail behavior of Gibbs free energy for $ P_\tau < P < P_{c2}$. Plots are displayed for $(Q, \alpha, \beta)=(1,0.01,0.38)$. } 
	}
\end{figure} 

\begin{figure}[h!]
	% \begin{figure}{l}{0.43\textwidth}
	%\begin{center}
	{\centering
		\includegraphics[width=3in]{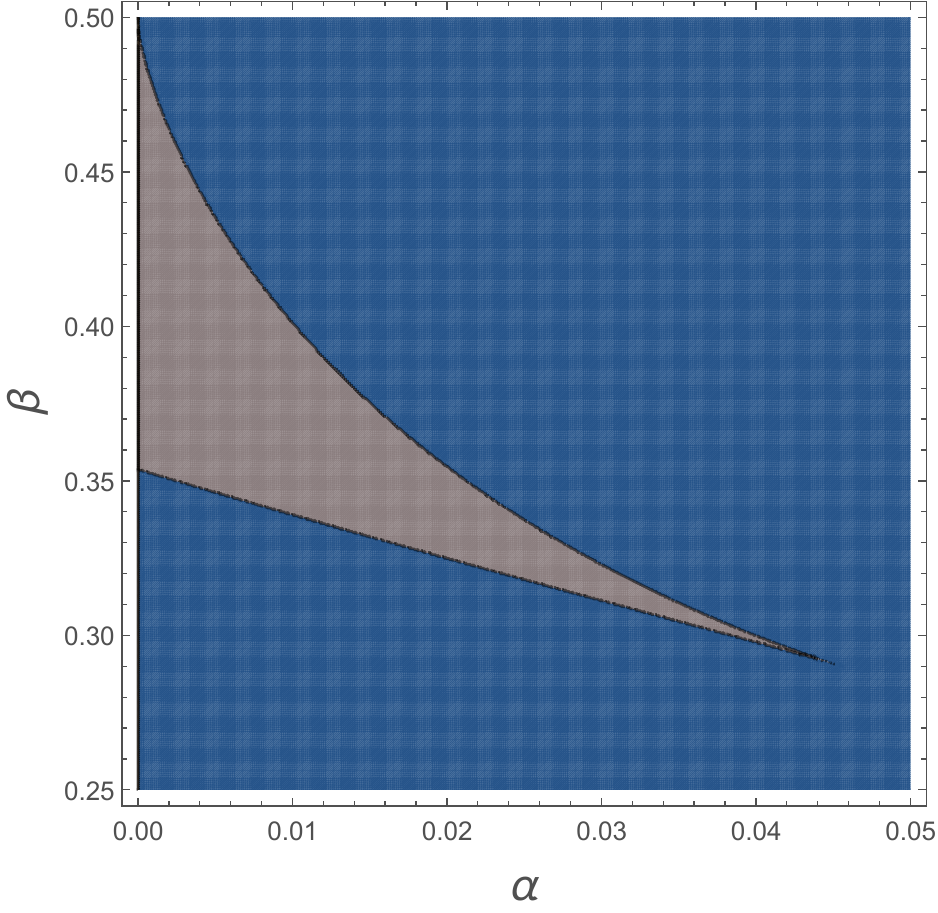}	
		%1		
		
		\caption{\footnotesize The available parameter region of $(\alpha, \beta)$, for the existence of small/intermediate/large black hole transitions and triple points, shown with gray colored region (where there can be three critical points). The blue colored region shows the existence of van der Waals type small/large black hole transitions (here there is only one critical point). Plot is displayed for $Q=1$.} \label{fig:alpha_beta_plot}
	}
\end{figure}
\noindent
Charged AdS black holes with Gauss-Bonnet coupling $\alpha$, in the novel four dimensional Einstein-Gauss-Bonnet gravity, are known to exhibit the phase structure of  van der Waals type small/large black hole first order transitions, terminating in a second order critical point~\cite{Wei:2020poh}.
On the other hand, with Born-Infeld coupling $\beta$, charged AdS black holes in four dimensions exhibit a phase structure containing reentrant phase transitions, along with the usual van der Waals type small/large black hole transitions~\cite{Gunasekaran:2012dq}.
\vskip 0.3cm \noindent
However, when  both the couplings $(\alpha, \beta)$ are present, the phase structure of charged AdS black holes gets an interesting modification. 
This can be seen as follows. For example, when $(Q, \alpha, \beta)=(1,0.01,0.26)$,  the equation of state  $P(r_+, T)$ (from eqn.~\ref{eq: temp}), shows the van der Waals fluid type behavior (see Fig.~\ref{Fig:pr1}). There exists a critical temperature $T_c$, below which there are small and large black hole branches, with a first order phase transition terminating at the second order critical point (see the Table~\ref{table:cri} for critical values). Below the critical point, the Gibbs free energy $G$ shows the swallow tail behavior (see Fig.~\ref{Fig:gt1}), which is the characteristic behavior of the first order transitions of a van der Waals fluid.
The coexistence line of first order small/large black hole transitions ending in a second order critical point is shown in the $P-T$ plot (see Fig.~\ref{Fig:pt1}). For $T>T_c$, black holes exist in a unique phase. \\
 \vskip 0.3cm
 \noindent
Now, consider another choice,  $(Q, \alpha, \beta)=(1,0.01,0.38)$. In this case, the equation of state exhibits an interesting structure as shown in the Fig.~\ref{Fig:pr3}). Now there exist three critical points $ \text{CP}_1$, $ \text{CP}_2$, and $\text{CP}_3$ (see the Table~\ref{table:cri} for critical values).
  The critical points $ \text{CP}_1$ and $ \text{CP}_2$, correspond to two first order small/intermediate and intermediate/large black hole transitions as shown in the Fig.~\ref{Fig:pt3}. The critical point $\text{CP}_3$ though does not correspond to a phase transition as  it does not  globally minimize the Gibbs free energy. Thus the critical point $\text{CP}_3$ is unphysical. 
  As shown in the Fig.~\ref{Fig:pt3}, all the three phases (i.e., the small, intermediate and large black hole phases) merge in a special point called triple point. This triple point in the $P-r_+$ diagram corresponds to a special isotherm for which a ‘double’ Maxwell equal area holds for the same pressure (see Fig.~\ref{Fig:prdst}).
  The Gibbs free energy shows a double swallow tail behavior, starting from the triple point to critical point $\text{CP}_2$,  corresponding to two first order transitions as shown in the Fig.~\ref{Fig:gtdst}. Thus, this phase structure of small/intermediate/large black hole phase transitions and the triple point, show resemblance to the solid/liquid/gas phase transition~\cite{AltamiranoKubiznak}.  
 \vskip 0.3cm
 \noindent
The available parameter region of $(\alpha, \beta)$, for the existence of small/intermediate/large black hole transitions and triple points~\footnote{All the critical points may not appear in the phase diagram. Only those appear that globally minimize the Gibbs free energy.}, is shown as gray colored region in Fig.~\ref{fig:alpha_beta_plot}. The van der Waals type small/large black hole transitions are shown as blue colored region in Fig.~\ref{fig:alpha_beta_plot}, where there is only one critical point.
The detailed discussion of phase structures related to these critical and triple points, can be found in~\cite{Zhang:2020obn}. 
\vskip 0.3cm \noindent
Our aim in this paper, is to study the topological properties associated with these critical points~\footnote{We do not consider the case of triple points here.} that belong to different parameter regions of $(\alpha, \beta)$ shown in the Fig.~\ref{fig:alpha_beta_plot}.

%%%%%%%%%%%%%%%%%%%%%%%%%%%%%%%%%
\subsection{Topological charge of critical points} \label{tcharge1}
%%%%%%%%%%%%%%%%%%%%%%%%%%%%%%%%%%%%%%%%
%Recently, it is shown in~\cite{Wei:2021vdx} that, the critical points  in the phase diagram of a  thermodynamic system (black holes in particular) could be classified into conventional and novel  critical type, based on the topological charges they carry. 
Let us outline the proposal in~\cite{Wei:2021vdx} for assigning topological charges to critical points. The temperature $T =  T(S, P, x^i) $ is typically a function of entropy $S$, pressure $P$ and other thermodynamic variables $x^i$, which can be extremised to get the critical points from
\begin{equation}\label{eq:inflection}
	(\partial_S T )_{P, x^i} =0, \ \ (\partial_{S,S} T )_{P, x^i} =0 \, .
\end{equation}
Above relation implies the possibility of constructing a Duan's potential~\cite{Wei:2021vdx}
\begin{equation}
	\Phi = \frac{1}{\text{sin} \theta} \, T(S, x^i) \, ,
\end{equation}
which in principle can be found upon eliminating one of the thermodynamic variables through eqn. (\ref{eq:inflection}), with a factor of $1/\sin\theta$ added for convenience. 
The set up of Duan's $\phi$-mapping theory can now be applied by constructing a vector field $\phi^a = (\phi^S, \phi^\theta)$, where $\phi^S = (\partial_S \Phi)_{\theta, x^i}$ and $\phi^\theta = (\partial_\theta \Phi)_{S, x^i}$.  The existence of a topological current $j^{\mu}$ satisfying $\partial_{\mu}\, j^{\mu}=0$ is a key feature, which gets non-zero contribution only from the points where the vector field $\phi^a$ vanishes, i.e., $\phi^a(x^i) = 0$. This ensures the existence of topological charge~\cite{Duan:1984ws,Duan:2018rbd,Wei:2021vdx}
\begin{equation}\label{tcharge}
 Q_t =\int_\Sigma j^0 d^2x= \Sigma_{i=1} w_i\, ,
\end{equation}
contained with in a region $\Sigma$ and $w_i$ is the winding number for the $i$-th point where $\phi$ is zero. Since, $Q_t$ can be positive or negative, the critical points were proposed to be further divided into two different topological classes, i.e., the conventional (where $Q_t = -1$) and the novel  (where $Q_t= +1$)~\cite{Wei:2021vdx}. If one allows $\Sigma$ to span the full parameter space of the thermodynamic system, the critical points of different thermodynamic systems can be classified into different topological classes.

%%%%%%%%%%%%%%%%%%%%%%%%
\section{Topology of critical points in  Born-Infeld AdS black holes in 4D Einstein-Gauss-Bonnet gravity} \label{4GBBI}
%%%%%%%%%%%%%%%%%%%%%%%%
First we obtain the general expression for topological charge which can be used to analyze the critical points of Born-Infeld AdS black holes in four dimensional Einstein-Gauss-Bonnet gravity.
%%%%%%%%%%%%%%%%%%%%%%%%
\subsection{Topological Charge} 
%%%%%%%%%%%%%%%%%%%%%%%%%%%%
Following the discussion in subsection-(\ref{tcharge1}), we start by computing the thermodynamic function $\Phi$, using the eqn~\eqref{eq: temp} for temperature~\footnote{where for a given value of $\alpha$, entropy $S=S(r_+)$ from eqn.~\eqref{eq: entropy}.} $T$, given by 
\begin{equation}
\Phi = \frac{1}{\text{sin}\theta} T(r_+, Q, \alpha, \beta) = \frac{1}{\text{sin}\theta} \frac{\bigg( r_+^2-2\alpha -\frac{2Q^2}{\sqrt{1+\frac{Q^2}{r_+^4\beta^2}}}\bigg)}{2\pi r_+ (r_+^2+6\alpha)}. \\
\end{equation}
\noindent
Then, the vector field $\phi= (\phi^r, \phi^\theta)$ is obtained to be:
\begin{eqnarray}
\phi^r & = &\partial_{r_+} \Phi \nonumber \\
&= &\frac{\text{csc}\theta}{2\pi (r_+^2+6\alpha)^2} \bigg\{  (r_+^2+6\alpha)\Big[ 2-\frac{4Q^4}{r_+^6\beta^2\Big( 1+\frac{Q^2}{r_+^4\beta^2}\Big)^{\frac{3}{2}}} \Big] - (3+\frac{6\alpha}{r_+^2}) \Big[ r_+^2-2\alpha -\frac{2Q^2}{\sqrt{1+\frac{Q^2}{r_+^4\beta^2}}} \Big]  \bigg\}, \nonumber \\
\phi^\theta &=& \partial_\theta \Phi =  -\frac{\text{cot}\theta \text{csc}\theta}{2\pi r_+ (r_+^2 +6\alpha)} \bigg( r_+^2-2\alpha -\frac{2Q^2}{\sqrt{1+\frac{Q^2}{r_+^4\beta^2}}} \bigg).
\end{eqnarray}
The normalised vector field can be computed from $n=(\frac{\phi^r}{||\phi||},\frac{\phi^\theta}{||\phi||})$.
The measurement of winding number $w_i$ enclosed by a given contour reveals the topological charge. If a critical point is contained inside a given contour, then its topological charge (i.e., the winding number $w_i$)  would be non-zero, otherwise the contour gives zero topological charge~\cite{Wei:2021vdx,Cunha:2020azh,Junior:2021svb,Wei:2020rbh,Cunha:2017qtt,Wei:2022mzv}.
\vskip 0.3cm \noindent
We consider a contour $C$,  which is piece-wise smooth and positively oriented in the orthogonal $\theta - r$ plane,  parameterized by the angle $\vartheta \in (0, 2\pi)$, as~\cite{Wei:2021vdx,Wei:2020rbh}:
\begin{eqnarray}
	\left\{
	\begin{aligned}
		r&=a\cos\vartheta+r_0, \\
		\theta&=b\sin\vartheta+\frac{\pi}{2}.
	\end{aligned}
	\right.
\end{eqnarray} \noindent
Then, we compute the deflection angle $\Omega({\vartheta})$ of the vector field $\phi$, along the contour $C$, given by
\begin{equation}
\Omega(\vartheta)=\int_{0}^{\vartheta}\epsilon_{ab}n^{a}\partial_{\vartheta}n^{b}d\vartheta,
\end{equation}
from which, one can obtain the topological charge $Q_t$, as
\begin{equation}
Q_t = w_i=\frac{1}{2\pi} \Omega (2\pi). \label{eq: qt_from_Omega}
\end{equation} \noindent
In the following subsections, we compute the topological charges corresponding to the critical points present in different parameter regions, obtained for various values of $(\alpha, \beta)$.
%%%%%%%%%%%%%%%%%%%%%%%%%%%%%%%
\subsection{\bf Case 1: One critical point}
\begin{figure}[h!]
	% \begin{wrapfigure}{r}{0.43\textwidth}
	%\begin{center}
	{\centering
		\subfloat[]{\includegraphics[width=3in]{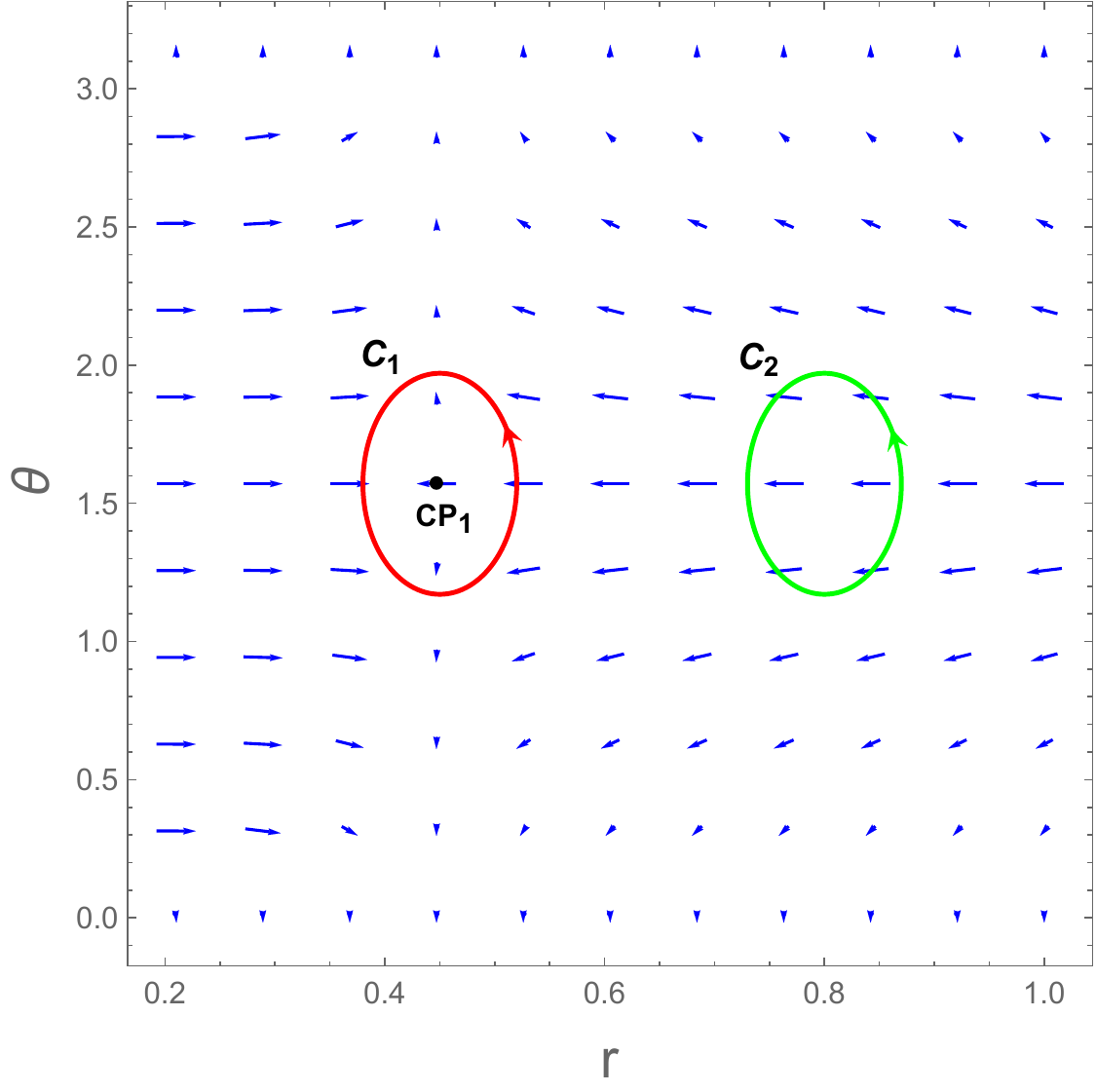}\label{Fig:vecplot_one_cp}}\hspace{0.5cm}	
		\subfloat[]{\includegraphics[width=3in]{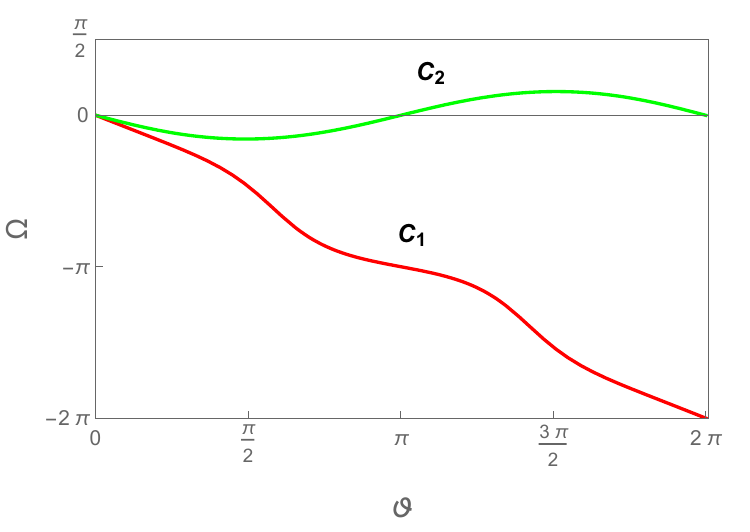}\label{Fig:omegaplot_one_cp}}				
		%5a,5b	
		
		\caption{\footnotesize For case-1: (a) The  vector field $n$ in $\theta-r$ plane, shows the presence of critical point $\text{CP}_1$ (black dot)  at $(r,\theta)=(0.446, \frac{\pi}{2})$. The  contour $C_1$ hides the critical point, while, the  contour $C_2$  does not. (b) $\Omega$ vs $\vartheta$ for contours $C_1$ (red curve), and $C_2$ (green curve).} 
	}
\end{figure}

%%%%%%%%%%%%%%%%%%%%%%%%%%%%%
This is the case where we have only one critical point (CP), associated with the van der Waals type small/large black holes transitions that occur in the blue colored region of Fig.~\ref{fig:alpha_beta_plot}. For this case, the vector field $n$, plotted in the Fig.~\ref{Fig:vecplot_one_cp}, shows the presence of the critical point (critical values are given in the Table~\ref{table:cri}). 
\begin{table}[!htb]
	\begin{minipage}{.5\linewidth}
		\centering{		 
			\begin{tabular}{|c|c|c|c|c|c|c|}
				\hline \hline
				\multicolumn{3}{|c|}{Case} & \multicolumn{3}{|c|}{Critical Point (CP)} \\   \cline{1-6} 
				& $\beta$ & & $\text{CP}_1$ & $\text{CP}_2$ & $\text{CP}_3$ \\ \hline
				\multirow{3}{1.2cm} {Case-1} & & $r_c$ & 0.4459 & - & -  \\ 
				& 0.26	& $P_c$ & 0.0374 & - & -  \\ 
				&	& $T_c$ & 0.1042 & - & - \\ \hline
				\multirow{3}{1.2cm} {Case-2} & & $r_c$ & 0.604 & 2.0845 & 1.0714 \\ 
				&	0.38	& $P_c$ & 0.00413  & 0.00373 & 0.00206 \\ 
				&		& $T_c$ & 0.0435  & 0.0453 & 0.0404 \\
				\hline\hline
			\end{tabular}
			\caption{Critical values at $Q=1$, $\alpha =0.01$.}
			\label{table:cri}	}
	\end{minipage} \hskip .2cm	
	\begin{minipage}{.5\linewidth}
		\centering{	
			\begin{tabular}{|c|c|c|c|c|c|c|}
				\hline \hline 
				\multicolumn{2}{|c|}{Case} &  $C_1$ & $C_2$ & $C_3$ & $C_4$  \\ \hline  
				\multirow{3}{1.2cm} {Case-1} & a & 0.07 & 0.07 & - & -  \\ 
				& b & 0.4 & 0.4 & - & -  \\ 
				& $r_0$ & 0.45 & 0.8 & - & -  \\ \hline
				\multirow{3}{1.2cm} {Case-2} & a & 0.15 & 0.15 & 0.15 & 1.1  \\ 
				& b & 0.4 & 0.4 & 0.4 & 0.7  \\ 
				& $r_0$ & 0.6 & 2.08 & 1.07 & 1.4  \\				
				\hline\hline 
			\end{tabular}
			\caption{Parametric coefficients of contours.}
			\label{table:coeff} }
	\end{minipage} 
	
\end{table}
\vskip 0.3cm \noindent
We construct two contours $C_1$ and $C_2$ for the vector field $n$ (parametric coefficients of the contours are given in the Table~\ref{table:coeff}). The contour $C_1$ contains the critical point, which gives non-zero contribution to the topological charge of the critical point, whereas, the contour $C_2$ does not contain the critical point, hence it gives zero topological charge. This can be seen from the behavior of the deflection angle $\Omega(\vartheta)$, along the contours $C_1$ and $C_2$, plotted in the Fig.~\ref{Fig:omegaplot_one_cp}. 
The angle $\Omega(\vartheta)$, along the contour $C_1$, decreases to reach $-2\pi$ at $\vartheta = 2\pi$, whereas, along the contour $C_2$, it first decreases, then increases and finally vanishes at $\vartheta =2\pi$.
\vskip 0.3cm \noindent 
Thus, in this case, the topological charge of the critical point $\text{CP}_1$ would be (from eqn.~\ref{eq: qt_from_Omega}) $Q_t|_{\text{CP}_1} = -1$, which is the total topological charge of the system as well (since, there is only one critical point).
%%%%%%%%%%%%%%%%%%%%%%%
\subsection{Case 2: \bf Three critical points } 
\begin{figure}[h!]
	% \begin{wrapfigure}{r}{0.43\textwidth}
	%\begin{center}
	{\centering
		\subfloat[]{\includegraphics[width=3in]{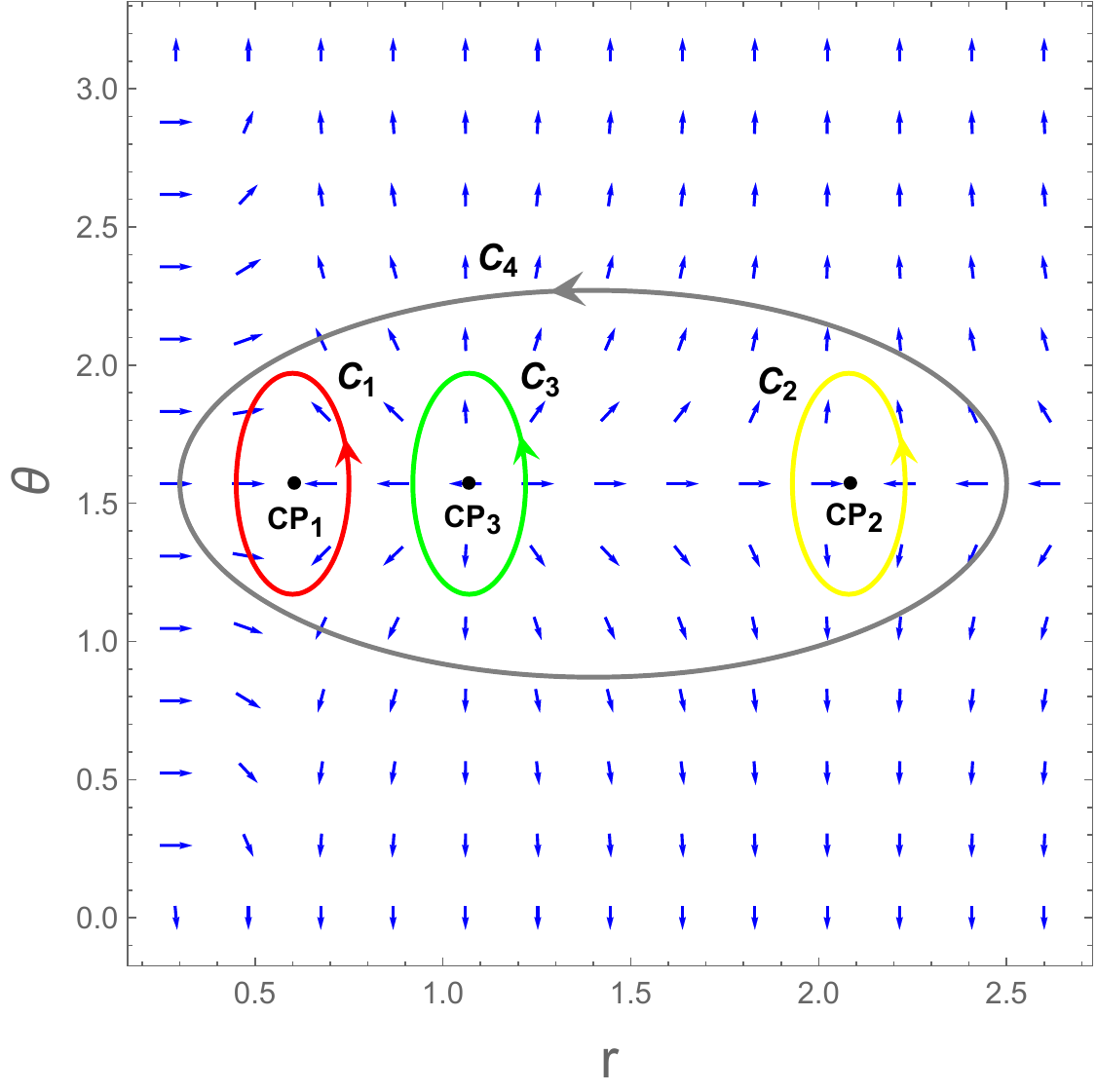}\label{Fig:vecplot_three_cp}}\hspace{0.5cm}	
		\subfloat[]{\includegraphics[width=3in]{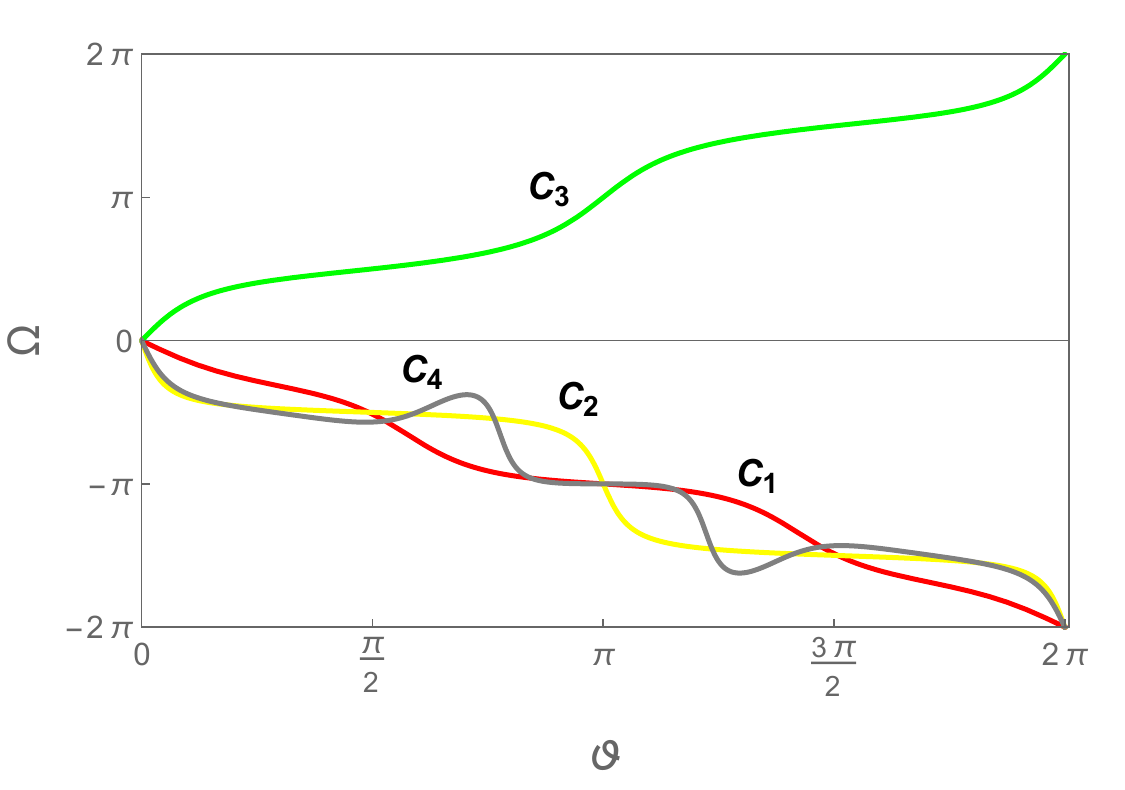}\label{Fig:omegaplot_three_cp}}				
		% 2a,2b	
		
		\caption{\footnotesize For case-2: (a) The  vector field $n$ in the $\theta-r$ plane, shows the presence of three critical points $\text{CP}_1$, $\text{CP}_2$, and $\text{CP}_3$ (black dots) at $(r,\theta)=(0.604, \frac{\pi}{2}), (2.085, \frac{\pi}{2}), \, \text{and} \, (1.07, \frac{\pi}{2}) $, which are contained inside the contours $C_1$, $C_2$ and $C_3$, respectively. 
			The  contour $C_4$   hides all the three critical points. (b) $\Omega$ vs $\vartheta$ for contours $C_1$ (red curve),  $C_2$ (yellow curve), $C_3$ (green curve), and $C_4$ (gray curve).} 
	}
\end{figure}

%%%%%%%%%%%%%%%%%%%%%%%%%
In this case, we have three critical points, associated  with the small/intermediate/large black hole transitions that occur in the gray colored region of Fig.~\ref{fig:alpha_beta_plot}. The presence of these three critical points (critical values are given in the Table~\ref{table:cri}) can be seen clearly from the vector field plot shown in Fig.~\ref{Fig:vecplot_three_cp}.
\vskip 0.3cm \noindent
For this case, we then consider four contours $C_1$, $C_2$, $C_3$, and $C_4$ (parametric coefficients of the contours are given in the Table~\ref{table:coeff}). 
The contours $C_1$, $C_2$,  and $C_3$, contain the critical points $\text{CP}_1$,  $\text{CP}_2$, and $\text{CP}_3$, respectively, while the contour $C_4$ contains all the three critical points.  
The behavior of the deflection angle $\Omega(\vartheta)$ for these contours is as shown in Fig~\ref{Fig:omegaplot_three_cp}, from which we find that $\Omega(2\pi) = -2\pi, -2\pi, +2\pi,  \, \text{and} \, -2\pi$, for the contours $C_1$, $C_2$, $C_3$, and $C_4$, respectively.
This results in the topological charges of the critical points coming out to be $Q_t|_{\text{CP}_1} = -1, \, Q_t|_{\text{CP}_2} = -1$, and $Q_t|_{\text{CP}_3} = +1$.
\vskip 0.3cm \noindent
Therefore, in this case, the total topological charge of the system (which is additive) would be again $-1$, given by the contour $C_4$.  
%%%%%%%%%%%%%%%%%%%%%%%
\subsection{\bf Limiting cases: } 
%%%%%%%%%%%%%%%%%%%%%%%%%
As mentioned earlier, in the limit of $\beta \rightarrow \infty$, we recover the charged AdS black holes in four dimensional Einstein-Gauss-Bonnet gravity, where we have only one critical point corresponding to the van der Waals type small/large black hole transitions. In this case, one can check that the topological charge of the critical point would be $-1$. 
On the other hand, in the limit of $\alpha \rightarrow 0$, we recover the Born-Infeld AdS black holes. In this case, we have two critical points corresponding to reentrant phase transitions for the range: $ \frac{1}{Q\sqrt{8}} < \beta < \frac{\sqrt{3+2\sqrt{3}}}{6Q}$, otherwise, there is only one critical point related to van der Waals type small/large black hole transitions~\cite{Gunasekaran:2012dq}. 
The topological charges of these critical points have been computed in~\cite{Wei:2021vdx}, where it has shown that the total topological charge of the system, in the case of two critical points, vanishes. 
%%%%%%%%%%%%%%%%%%%%%%%%%%%%%%
\subsection{Nature of the Critical points} \label{naturegbbi}
\begin{figure}[h!]
	
	% \begin{wrapfigure}{r}{0.43\textwidth}
	%\begin{center}
	{\centering
		
		\subfloat[]{\includegraphics[width=2.9in]{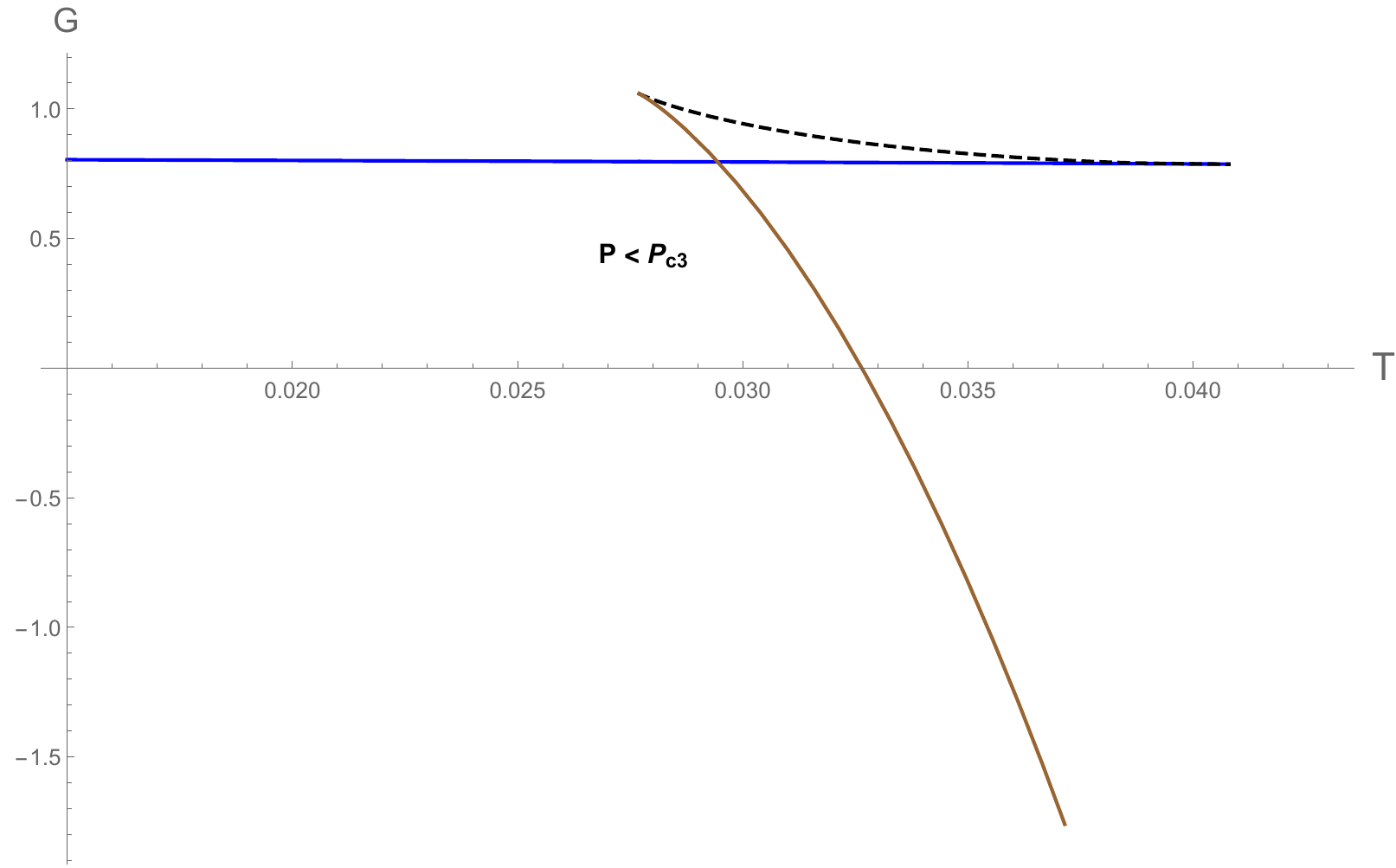}}\hspace{0.5cm}	
		\subfloat[]{\includegraphics[width=2.9in]{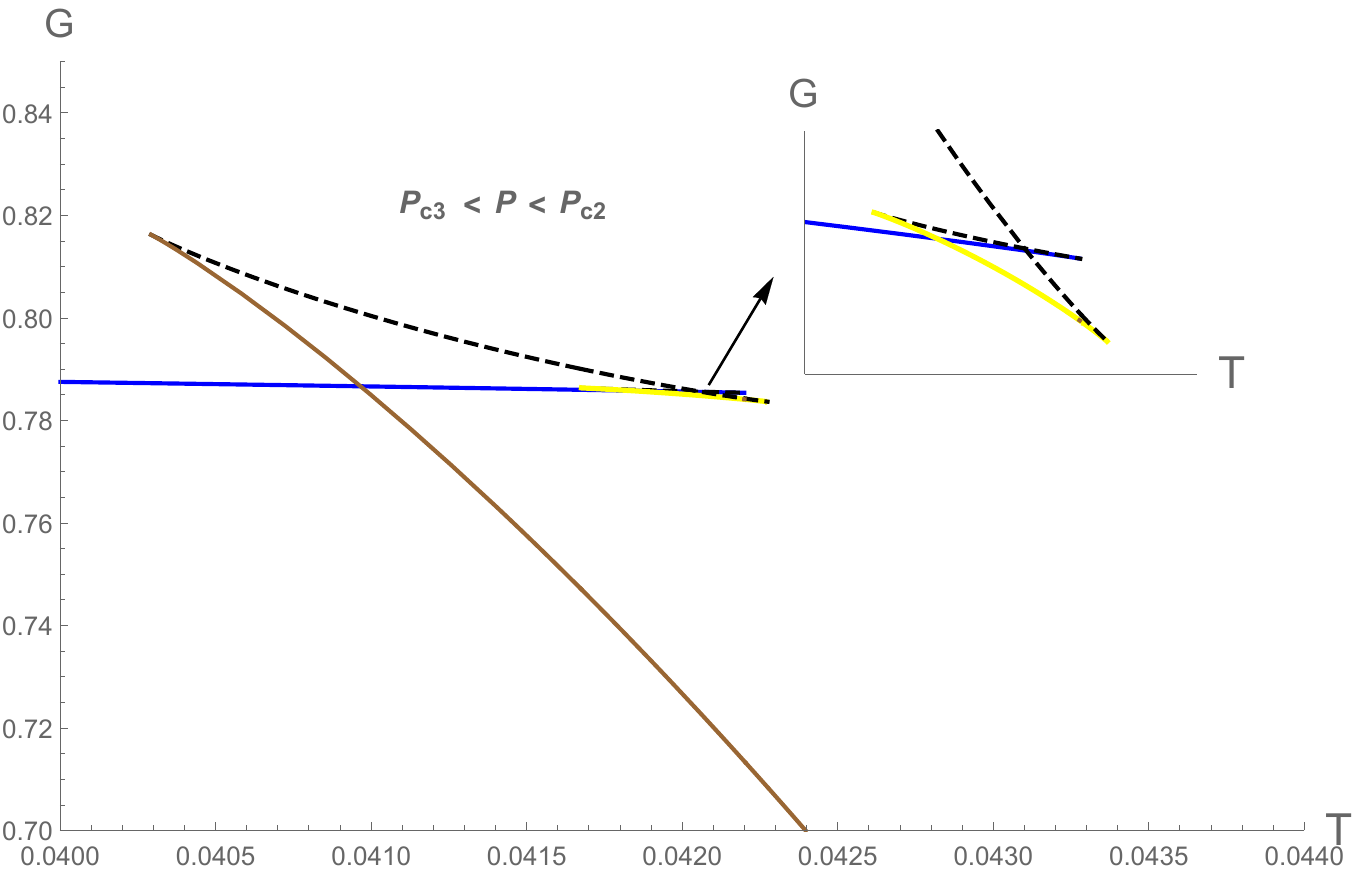}}\hspace{0.5cm}				
		\subfloat[]{\includegraphics[width=2.9in]{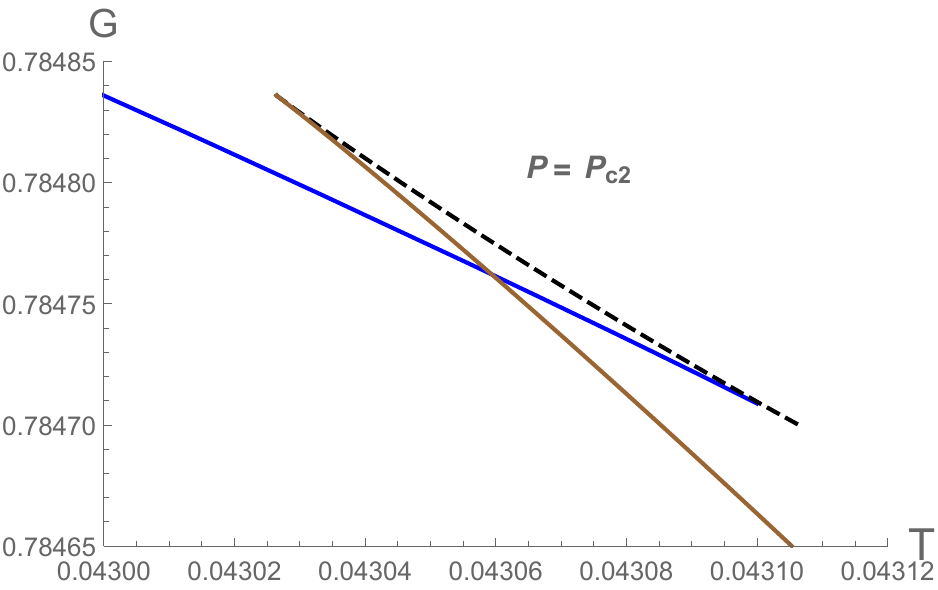}}\hspace{0.5cm}				
		\subfloat[]{\includegraphics[width=2.9in]{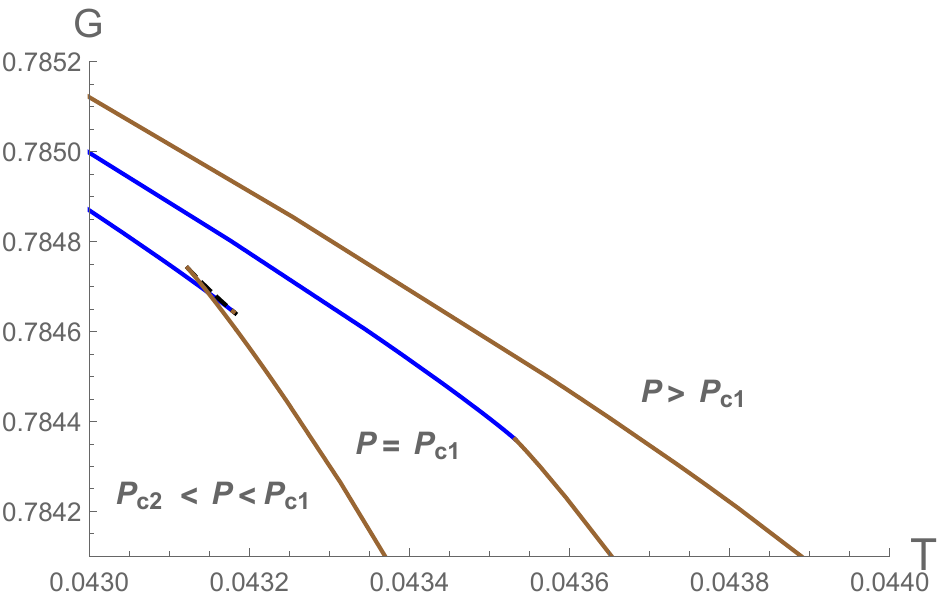}}
		
		%7a,7b,7c,7d	
		
		\caption{\footnotesize For case-2: Behaviour of the Gibbs free energy as a function $T$ for different pressures,  showing  the  appearance of new (stable or unstable) phases near the novel critical point $\text{CP}_3$ (with pressure $P_{c3}$), and the disappearance of  phases near the conventional critical points $\text{CP}_2$ ( with pressure $P_{c2}$), and $\text{CP}_1$ (with pressure $P_{c1}$),   on increase of pressure. (a) $P=0.0013$, (b) $P=0.0028$,  (c) $P=P_{c2} = 0.0037$,  (d) $P=\{0.0038, P_{c1}=0.0041, 0.0045\}$.} 
		\label{fig:gtplots_three_cp}	}
	
\end{figure}

%%%%%%%%%%%%%%%%%%%%%%%%%%%%%%%
From the view of topology, critical points have been classified into conventional ($Q_t = -1$) or novel ($Q_t = +1$), based on the topological charge they carry~\cite{Wei:2021vdx}. 
A further proposal of~\cite{Yerra:2022alz} is that, in the phase diagram as the pressure increases, conventional critical point acts as the phase annihilation point, where the phases  disappear, whereas, the novel critical point acts as the phase creation point, where the new (stable or unstable) phases appear. This classification scheme continues to hold for isolated critical points~\cite{Ahmed:2022kyv} and in the current case too, as we summarise below.
\begin{figure}[h!]
	% \begin{wrapfigure}{r}{0.43\textwidth}
	%\begin{center}
	{\centering
		\subfloat[]{\includegraphics[width=2.9in]{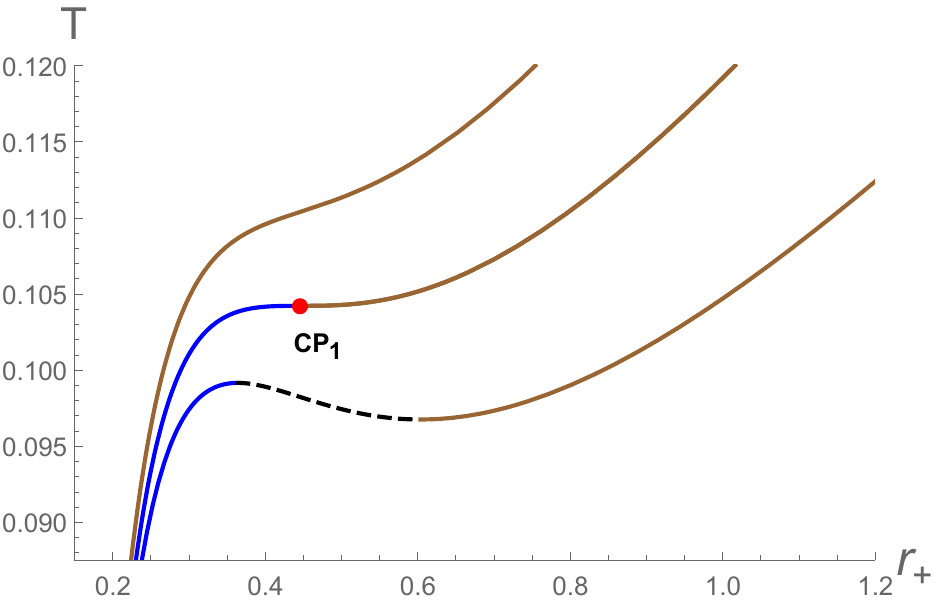}\label{Fig:trplot_one_cp}}\hspace{0.5cm}	
		\subfloat[]{\includegraphics[width=2.9in]{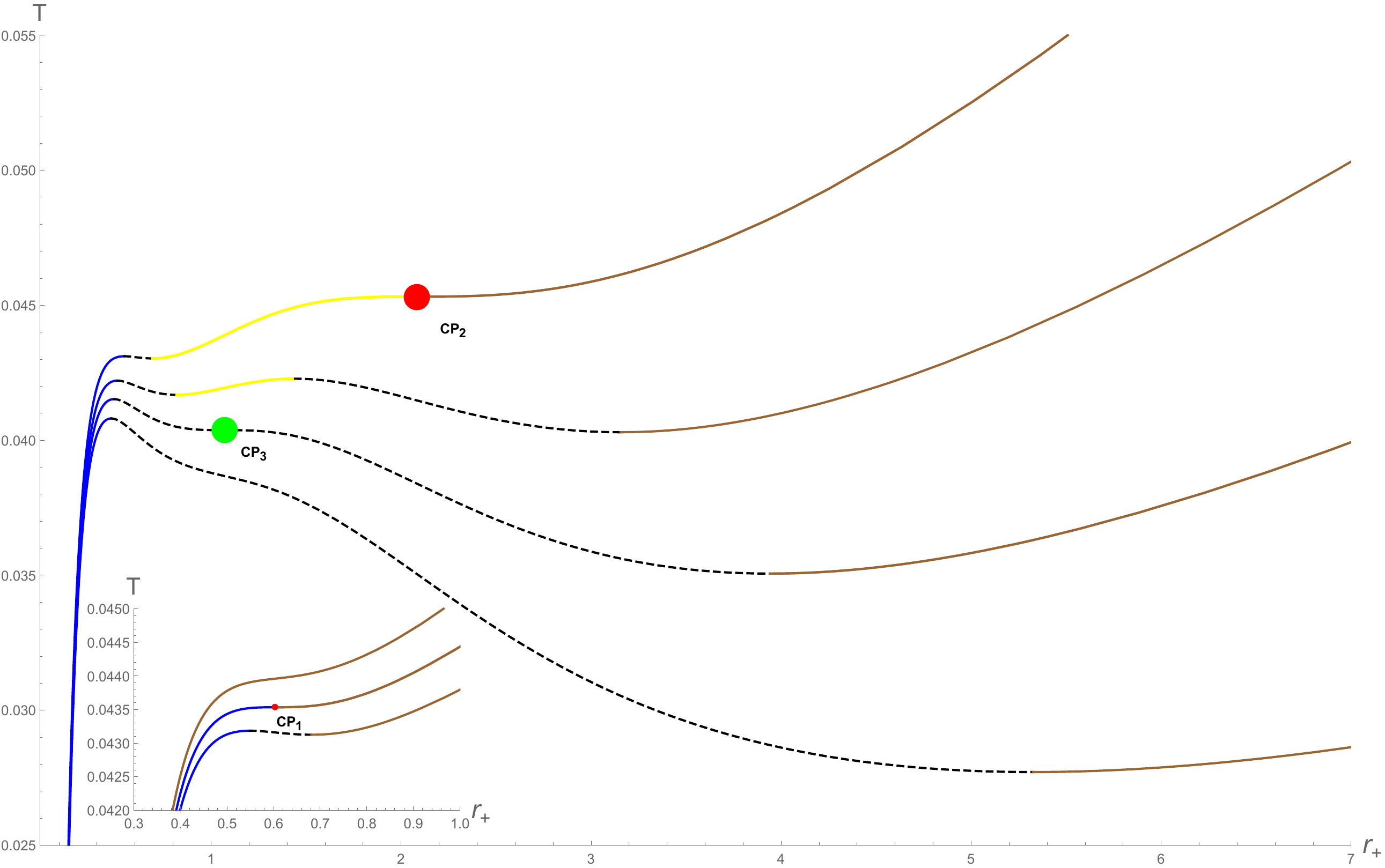}\label{Fig:trplot_three_cp}}				
		%8a,8b	
		
		\caption{\footnotesize   $T$ as a function of $r_+$ for different pressures, showing the appearance of new (stable or unstable) phases near the novel critical point (green dot), and disappearance of phases near the conventional critical point (red dot) as the pressure increases.  Dashed curves denote unstable black hole branches and solid curves denote stable black hole branches. Pressure of the isobars increases from bottom to top. (a) for case-1, (b) for case-2 (Inset: for  $P > P_{c2}$).} 
	}
\end{figure}  
\vskip 0.1cm \noindent
For case-1 discussed earlier, there was one critical point $\text{CP}_1$, for which the topological charge is $Q_t = -1$ (conventional). 
The phase structure around this critical point is as shown in the Fig.\ref{Fig:trplot_one_cp}, where the number of phases in an isobar decreases as the pressure increases, making it a phase annihilation point.
For case-2, there were two conventional critical points $( \text{CP}_1\, \text{and} \, \text{CP}_2)$, and one novel critical point $(\text{CP}_3)$. As the pressure increases, the disappearance of phases at the conventional critical points and the appearance of new (stable or unstable) phases at novel critical point, can be seen from the Fig.~\ref{Fig:trplot_three_cp}.
The behaviour of  Gibbs free energy $G$ around these critical points, plotted in Fig.~\ref{fig:gtplots_three_cp}, clearly shows the corresponding phase creation and annihilation behaviours.  
%%%%%%%%%%%%%%%%%%%%%%%%%%%%%%%%%%%%%%%%%%%
\section{Conclusions} \label{conclusions}
%%%%%%%%%%%%%%%%%%%%%%%%%%%%%%%%%%%%%%%%%%%%% 
We investigated the topological classification of critical points in the Born-Infeld AdS black holes in four dimensional novel Einstein-Gauss-Bonnet gravity in the extended phase space. With only the Born-Infeld corrections to AdS black holes, there are just reentrant phase transitions, whereas with the addition of Gauss-Bonnet coupling $\alpha$, the system exhibits small/intermediate/large black hole transitions and triple points too.
\vskip 0.3cm \noindent
Depending on the range of Gauss-Bonnet and Born-Infeld couplings $(\alpha, \beta)$, the phase structure of these black holes can admit one or three critical points, related to usual van der Waals type small/large black hole transitions or an interesting small/intermediate/large black hole transitions, respectively. We computed the topological charges associated with these critical points and the results are summarized in Table~\ref{table:summary}.
\begin{table}[h!]
	\centering{\
		\begin{tabular}{|c |c |c|c|c |c|c|}
			\hline \hline 
			Case & Number of  & Topological & Total Topological \\
			&  Critical Points &  Charge $Q_t$ &  Charge  \\ \hline
			\multirow{3}{11em}{Case 1:} &  &  &   \\ 
			& 1 & $Q_t|_{\text{CP}_1} = -1$ & $-1$  \\ 
			&  &  &   \\ \hline
			\multirow{3}{11em}{Case 2:} &  & $Q_t|_{\text{CP}_1} = -1$ &   \\ 
			& 3 & $Q_t|_{\text{CP}_2} = -1$ & $-1$  \\ 
			&  & $Q_t|_{\text{CP}_3} = +1$ &   \\ \hline
			\multirow{2}{11em}{Limiting Case: $ \alpha \rightarrow 0 $~\cite{Wei:2021vdx}} &  &  $Q_t|_{\text{CP}_1} = +1$ &   \\ 
			& 2 & $Q_t|_{\text{CP}_2} = -1$  & 0  \\ \hline
			\multirow{3}{11em}{Limiting Case:  $ \beta \rightarrow \infty $ } &  &  &   \\ 
			& 1 & $Q_t|_{\text{CP}_1} = -1$ & $-1$  \\ 
			&  &  &   \\ \hline			
		\end{tabular}
		\caption{Summary of results}
		\label{table:summary}}
\end{table}
\vskip 0.3cm \noindent
It was noted earlier that the Born-infeld corrections to the Einstein action alter the topological class of the charged AdS black hole system~\citealp{Wei:2021vdx}, whereas the Gauss-Bonnet, and more generally the Lovelock type corrections do not~\citealp{Yerra:2022alz}. Here, we showed that when topology of critical points is studied with both the above Gauss-Bonnet and Born-Infeld  corrections side by side (in four dimensions), the topological nature is unaltered, possibly indicating the dominance of former set of terms. Further, our results also obey the parity conjecture of critical points proposed in~\cite{Yerra:2022alz}, which says that, ``for odd (even) number of critical points, the total topological charge is an odd (even) number".  \\

\noindent
Relying on the set of examples studied thus far it would be nice to extend our conclusions for situations where there are general higher derivative gauge/gravity corrections to Einstein action. An important point to check is whether the nature of higher derivative gravity corrections to Einstein action is to preserve the topological class of black hole critical points, whereas that of gauge corrections is to alter it.  The competing nature for changing the topological charge of critical points, between the Gauss-Bonnet and Born-Infeld corrections in higher dimensions and with the possible inclusion of other types of non-linear gauge/gravity terms with multiple critical points should be analysed too~\cite{Sherkatghanad:2014hda,Frassino:2014pha,Sherkatghanad:2014hda,Ye:2022uuj,Astefanesei:2021vcp,Dykaar:2017mba,Tavakoli:2022kmo}.

%%%%%%%%%%%%%%%%%%%%%%%%%%%%
\section*{Acknowledgements}
One of us (C.B.) thanks the DST (SERB), Government of India, for financial support through the Mathematical Research Impact Centric Support (MATRICS) grant no. MTR/2020/000135. We thank the referee for critical review and suggestions which improved the manuscript.
%%%%%%%%%%%%%%%%%%%%%%%
\bibliographystyle{apsrev4-1}
\bibliography{topology_4dgbbi}
\end{document}